\documentclass[epj]{webofc}

\usepackage[varg]{txfonts}   
\usepackage{ upgreek }

\wocname{EPJ Web of Conferences}
\woctitle{ICNFP 2017}
\begin{document}
\selectlanguage{english}
\title{Ongoing magnetic monopole searches with IceCube}
%
%

\author{Frederik Lauber \thanks{\email{lauber@uni-wuppertal.de}} \inst{1} on behalf of the IceCube Collaboration \thanks{\url{https://icecube.wisc.edu}}
}

\institute{
Dept. of Physics, University of Wuppertal, 42119 Wuppertal, Germany
}

\abstract{%
The IceCube collaboration has instrumented a cubic kilometer of ice with $5160$ photo-multipliers.
While mainly developed to detect Cherenkov light, any visible light can be used to detect particles within the ice.\\
Magnetic monopoles are hypothetical particles predicted by many theories that extend the Standard model of Particle Physics.
They are carriers of a single elementary magnetic charge.
For this particle, different light production mechanisms dominate from direct Cherenkov light at highly relativistic velocities $\left(>0.76\,c\right)$, indirect Cherenkov light at mildly relativistic velocities $\left(\approx 0.5\,c \textrm{ to } 0.76\,c\right)$, luminescence light at low relativistic velocities $\left( \gtrsim 0.1\,c \textrm{ to } 0.5\,c\right)$, as well as catalysis of proton decay at non relativistic velocities $\left(\lesssim 0.1\,c\right)$.\\
For each of this speed ranges, searches for magnetic monopoles at the IceCube experiment are either in progress or they have already set the world's best limits on the flux of magnetic monopoles.
A summary of these searches will be presented, outlining already existing results as well as methods used by the currently conducted searches.
}
\maketitle

\section{Introduction}
\label{intro}
\begin{figure}[h]
\centering
\includegraphics[width=\textwidth,clip]{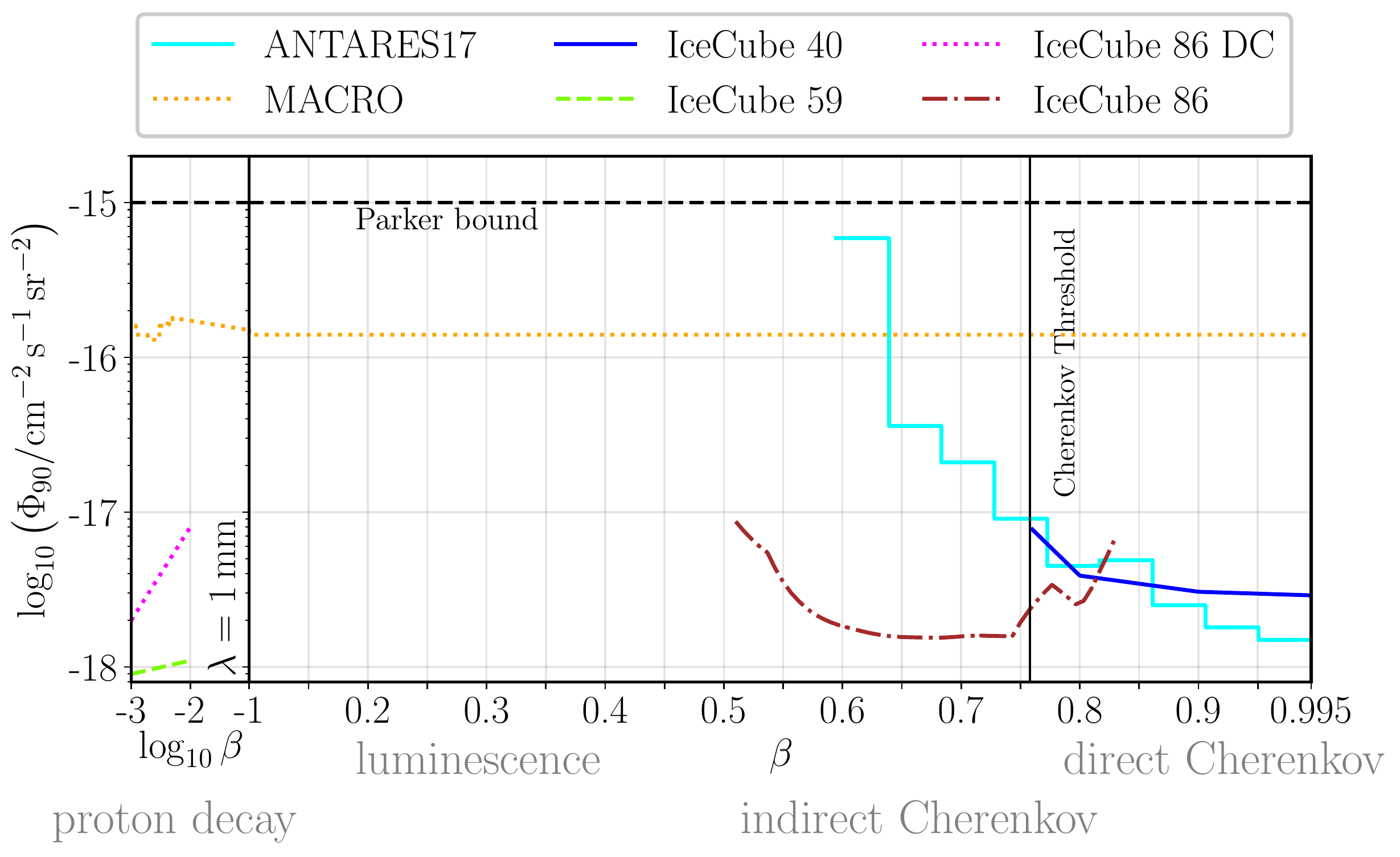}
\caption{Upper flux limits for magnetic monopoles as a function of their speed. Annotations for the major light production mechanism in the speed range are shown. Only current best limits are highlighted~\cite{hep-ex/0207020v2, 1402.3460v2, 1511.01350v2, Antares17}.
}
\label{fig-1}       
\end{figure}

Since Maxwell published his equations, describing the relation between the electric and magnetic fields in 1864,
the reason for the asymmetry between the magnetic and the electric field has been an open issue. Only with the introduction of a magnetic counterpart to the electron, the magnetic monopole, do the Maxwell equations become symmetric. The question of the existence of this hypothetical particle has become more important as all Grand Unifying Theories (GUTs) as well supersymmetry (SUSY) theories predict them~\cite{1707.08067v2}.\\ 
Multiple searches for magnetic monopoles have been conducted at multiple experiments over the past several years. The minimal magnetic charge of a magnetic monopole is known from theory~\cite{Dirac60} but there is no intrinsic limit on the mass such a particle would possess. Searches range from $10^{17}\,\textrm{GeV}$ (relic monopoles) to $10^{13}\,\textrm{GeV}$ (intermediate mass monopoles) down to $10^{3}\,\textrm{GeV}$ (light monopoles). This puts the lowest mass 
magnetic monopoles already at a potential detection level in current accelerator experiments, while the heavier ones will only be detectable as remnants of the big bang. 
\\
To detect such rare remnants, a high effective area is needed which makes IceCube well-suited for these searches.
The current flux limits as a function of the speed of the magnetic monopole are shown in Fig.~\ref{fig-1}.

\section{IceCube}
\label{sec-icecube}
The IceCube South Pole Neutrino Observatory is a neutrino detector situated at the geographical south pole~\cite{1612.05093v2}. It consists of $5160$\, Digital Optical Modules (DOMs) which are submerged into $1\,\textrm{km}^3$ of Antarctic ice. Each DOM consists of one downward looking photo-multiplier and supporting electronics inside a glass pressure sphere. The DOMs are sensitive to light in the $350\,\textrm{nm}$ to $650\,\textrm{nm}$ range. The dark rate of each DOM is approximately $480\,\textrm{Hz}$~\cite{1002.2442v1}.\\
The DOMs are organized in $86$\,strings in a roughly hexagonal grid. 
All data and power is transmitted to and from each DOM via the string to which it is connected.\\
Any light recorded by a DOM is separated into two classes of hits, Hard Local Coincidence (HLC) or Soft Local Coincidence (SLC). Any hit, where at least the nearest or next-nearest neighbor DOM on the string also recorded light, is an HLC hit, any other hit is a SLC hit. 
The full record of an HLC hit is transmitted to the surface. At the surface, a trigger may be generated which causes a complete readout of all HLC and SLC hits.
The recorded data is stored at the surface. After passing further filters, the data is transmitted via satellite to data centers in the north and then made available to the collaboration for analysis.

\section{Light production in ice}
\label{sec-prodlight}
\begin{figure}[h]
\centering
\includegraphics[width=\textwidth,clip]{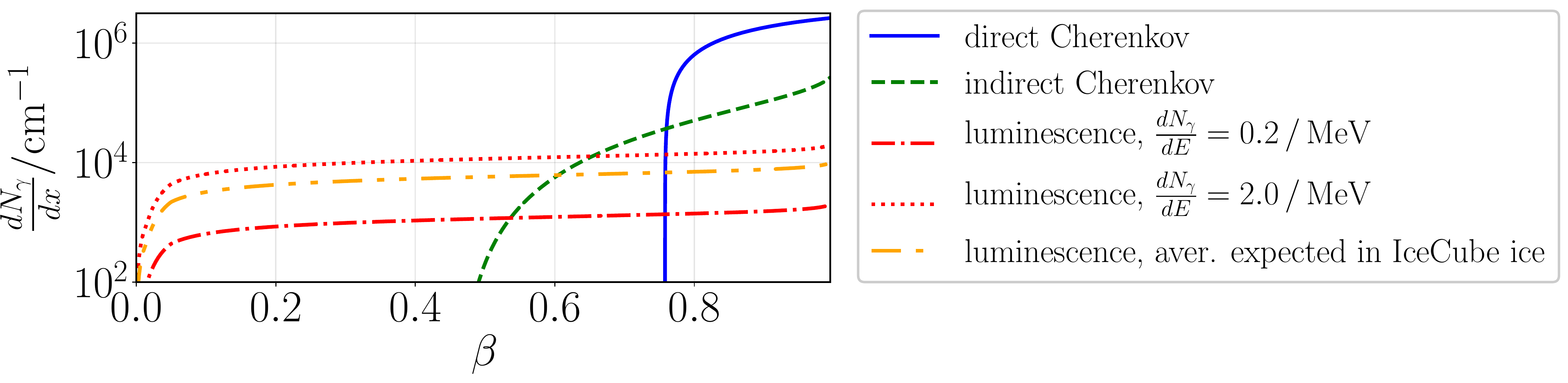}
\caption{Light yield for different light production mechanisms of magnetic monopoles in ice.}
\label{fig-2}       
\end{figure}

Magnetic monopoles are highly ionizing particles.
As such, multiple light production mechanisms in a medium exist.
For speeds above $0.76\,\textrm{c}$, Cherenkov light 
is produced in ice. Falling below the Cherenkov limit, 
the light production is dominated by Cherenkov light from $\delta$-electrons which are knocked out of
their respective atoms due to the monopole passing through the medium. This is called indirect Cherenkov light in contrast to direct Cherenkov light from the monopole itself. 
While there is no lower threshold for this light production mechanism,
the light yield drops rapidly below $0.6\,\textrm{c}$. Below $0.5\,\textrm{c}$, the light emission is dominated by luminescence light. This is produced by the excitation of the medium due to the energy deposition of an incident particle. Multiple measurements exist for the light yield of luminescence, covering roughly the range between $0.2\,\gamma/\textrm{MeV}$ to $2\,\gamma/\textrm{MeV}$. New measurements for the light yield of luminescence in IceCube ice are ongoing and measured a light yield of $1\,\gamma/\textrm{MeV}$ on average for IceCube's temperatures integrated from $250\,\textrm{nm}$ to $650\,\textrm{nm}$~\cite{ICRC17}. The expected light yields for all described light production mechanisms are displayed in Fig.~\ref{fig-2}. \\
In addition to the aforementioned mechanisms, there is
a monopole-specific production mechanism proposed to exist: proton catalysis. In
the generic model used in IceCube analyses, the monopole catalyzes protons to decay into a positron and multiple secondary particles. These in turn produce light. To be detectable in IceCube, the mean free path between subsequent decays needs to be below about $10\,\textrm{m}$.

\section{Current searches at IceCube}
\label{sec-current}
All analyses at the IceCube South Pole Neutrino Observatory are performed in a \textit{blind} manner by only using $10\,\%$ of the acquired data as well as simulated data for signal and background signatures. 
The remaining $90\,\%$ of the acquired data is only looked at after all selection criteria have been finalized.\\
Three new searches for magnetic monopoles are in progress. The searches are split into different speed regions of the incident particle which coincides with the different light production mechanisms mentioned above. 
The analyses and methods are discussed below.

\subsection{Highly relativistic Monopoles}
\label{sec-highrel}
\begin{figure}[h]
\centering
\sidecaption
\includegraphics[width=0.4\textwidth,clip]{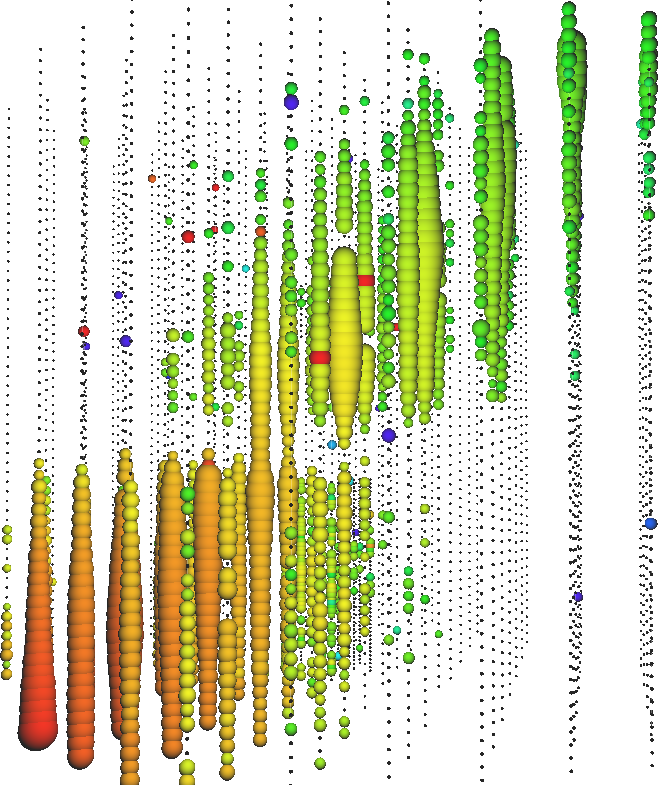}
\caption{Simulated signal event of an magnetic monopole at $0.982\,\textrm{c}$ traveling from the bottom to the top of the detector (up going). The color, from red to blue, indicates the time evolution of the event. The total time of the event is $5000\,\textrm{ns}$. The size of the spheres scales with the amount of light detected by each DOM. Few noise hits contribute due to the short time frame of the event. Less light has been detected in a horizontal plane roughly at the half height of the detector due to dust in the ice~\cite{1612.05093v2}.}
\label{fig-3}
\end{figure}

Highly relativistic magnetic monopoles are defined to have a speed  
above $0.76\,\textrm{c}$. An example simulated signal event is shown in Fig.~\ref{fig-3}. The events are bright and of short duration ($\mathcal{O}\left(10^3\,\textrm{ns}\right)$). \\
For the first time, the extremely high energy (EHE) neutrino sample~\cite{1304.5356v2} ($\mathcal{O}\left(1\,\textrm{event}/\textrm{year}\right)$) is used to detect monpoles. Additional cuts are then developed using simulated signatures of background and signal to separate highly energetic leptons produced by neutrino interactions inside the detector from magnetic monopoles. As these leptons are produced inside the detector volume and, in the case of the electron and tau, get stopped inside the detector volume, these events are called starting and stopping events.   
At the moment, the deposition of light along the direction of travel in correlation to the detector geometry
is investigated to remove starting and stopping events. As magnetic monopoles are not produced inside the detector nor do they get stopped inside the detector, this feature can be used to separate them from background.

\subsection{Low relativistic Monopoles}
\label{sec-lowrel}
\begin{figure}[h]
\centering
\sidecaption
\includegraphics[width=0.4\textwidth,clip]{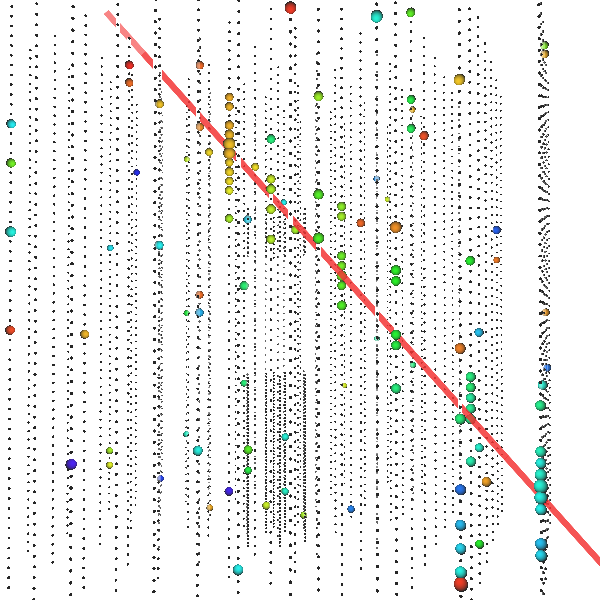}
\caption{Simulated signal event of a magnetic monopole with speed $0.3\,\textrm{c}$ moving from the top of the detector to the bottom (down going). The color, from red to blue, indicates the time evolution of the event. The total duration of the event is $30\,\textrm{$\upmu$s}$. The size of the spheres scales with the amount of light detected by each DOM. The original track of the particle was included in red.\vfill
The hits far away from the track are noise hits.}
\label{fig-4}
\end{figure}

\begin{figure}[h]
\centering
\sidecaption
\includegraphics[width=0.49\textwidth,clip]{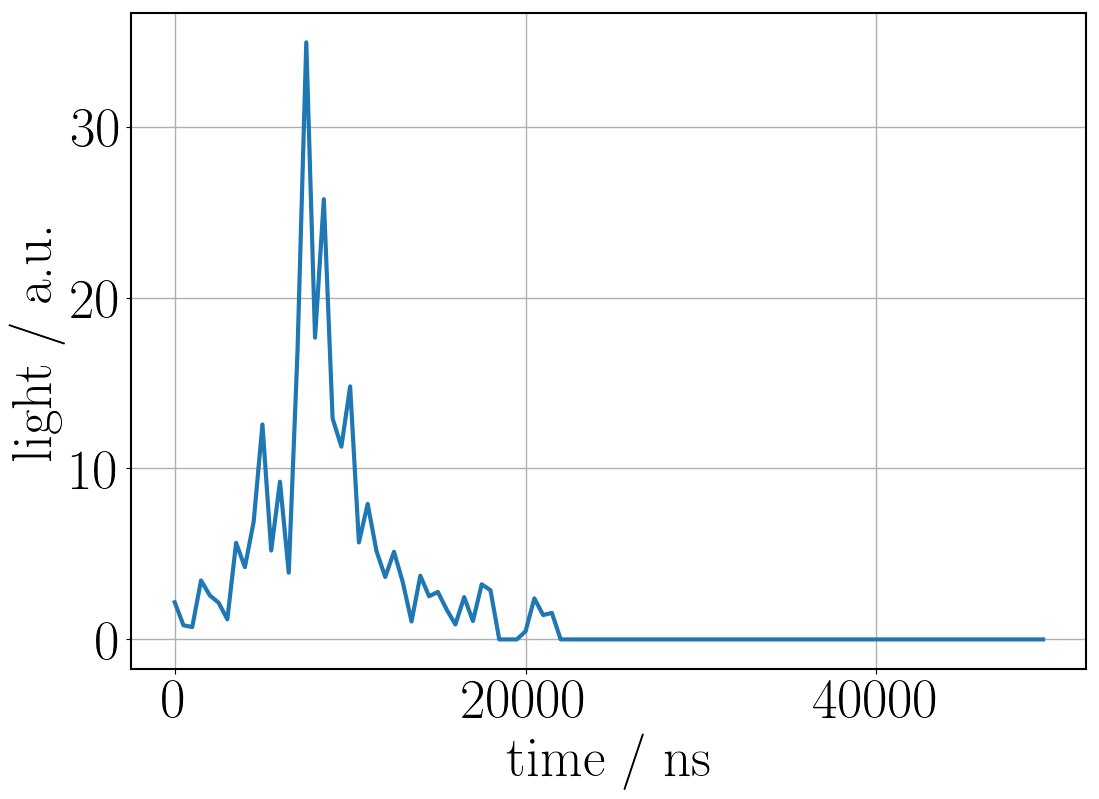}
\caption{Spatially compressed light curve of a simulated magnetic monopole event. All light recorded by all DOMs is included and shown as a function of time.}
\label{fig-5}
\end{figure}

\begin{figure}[h]
\centering
\includegraphics[width=0.49\textwidth,clip]{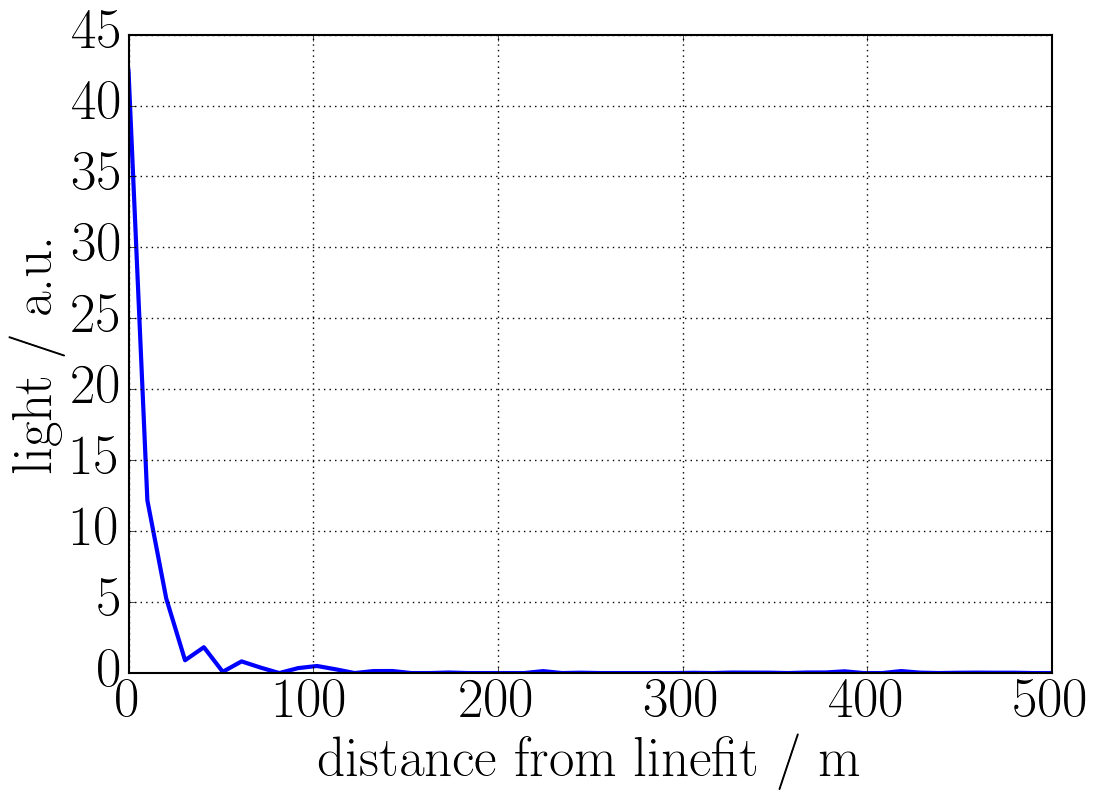}
\includegraphics[width=0.49\textwidth,clip]{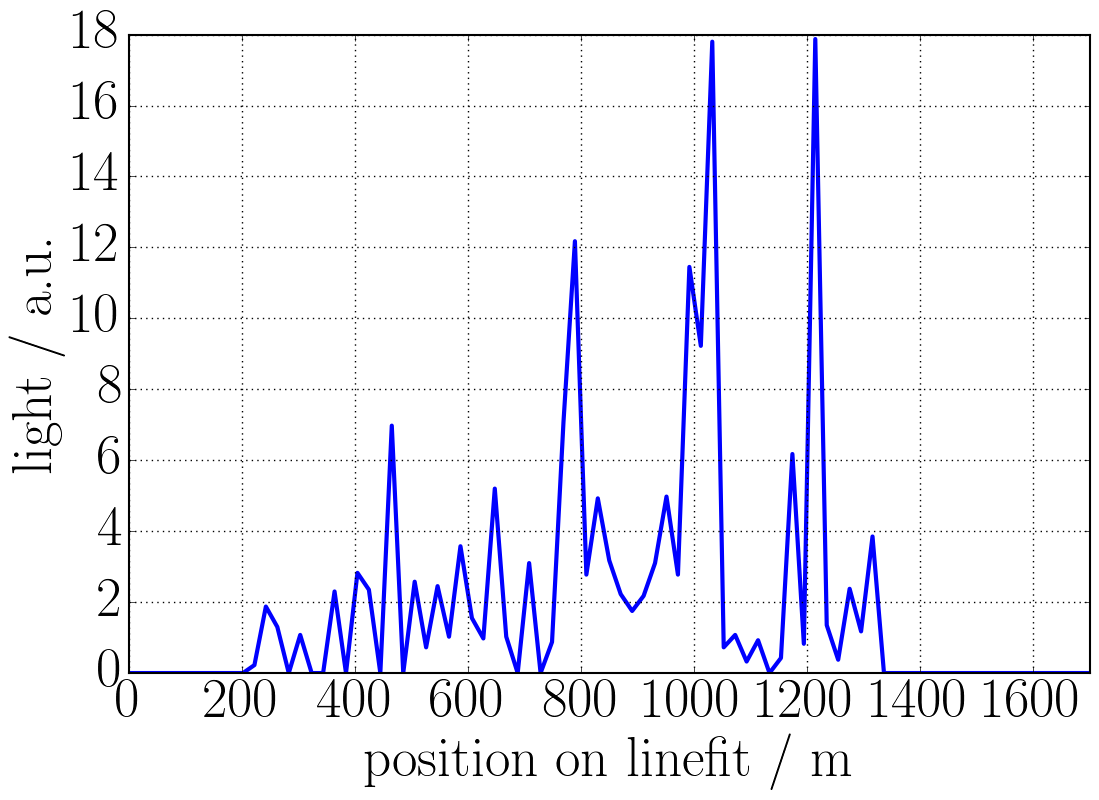}
\caption{Light of a simulated monopole event projected on the track hypothesis. On the left, light distribution as a function of to the radial distance from the track hypothesis. On the right, light deposited along the track hypothesis.}
\label{fig-track-hypothesis-projection}
\end{figure}

Low relativistic monopoles are defined to have a speed of 
$0.1\,\textrm{c}$ to $0.5\,\textrm{c}$. An example simulated signal event is displayed in Fig.~\ref{fig-4}. In contrast to muon and neutrino signatures
seen by IceCube, they are relatively dim. A special filter, the Monopole filter~\cite{1610.06397v1}, is in place in IceCube to select these events; it selects candidate events at a rate of $31\,\textrm{Hz}$. The filter works by placing straight cuts on different variables like the reconstructed velocity, reconstruction quality, and event duration.\\
Due to the low expected flux of magnetic monopoles based on current limits as seen in Fig.~\ref{fig-1}, the integrated expected signal rate for the IceCube detector is below $10^{-6}\,\textrm{Hz}$.
\newpage
As a first step, a linear regression of all hits is computed and a cut on the reconstructed
speed applied. The remaining events in the simulated and actual data sample are used to train a neural network~\cite{schmidhuber2015learning}.
The challenge here is twofold: encoding the physical information in a sensible way and defining a neural network structure which fits the problem. Both these problems are heavily constrained by processing power. Removing all spatial information a \textit{spatially compressed} light curve is created as seen in Fig.~\ref{fig-5}.
A first neural network trained on this spatial compressed light curve shows a separation power of two orders of magnitude between signal and background signatures. Further encoding is experimented with, in particular the projection of the produced light on the aforementioned track hypothesis (see Fig.~\ref{fig-track-hypothesis-projection}).\\
Neural networks can be expressed by chaining matrix operations with activation functions to link a given input to a specific output as seen in Eq.~\ref{equ-1}:
\begin{equation}
\label{equ-1}
\textrm{Output} = f_N\left(w_N*\dots f_0\left(w_0*\textrm{Input}\right)\right)
\end{equation}  
By feeding the neural network with examples of an input with its respective output, the weight matrices $w_N$ are adjusted so that the response of the network to the input matches the required output.
While there is no intrinsic limit on the form of the activation function, the presented analysis limits itself to functions which have an output with the same dimensions as the input (no convolution)
as the input is assumed to have low correlation between neighboring data points.
In this analysis, the number of layers $N$ is below $4$ to limit processing time during training.
As all of the aforementioned encodings are one dimensional, a fully connected neural network is utilized.
This means that the weight matrix $w_N$ is not sparse. The output is always a two dimensional vector containing the probabilities of the given input event to be either a background or monopole signature.

\subsection{Slow Monopoles}
\label{sec-slowrel}
\begin{figure}[h]
\centering
\sidecaption
\includegraphics[width=0.4\textwidth,clip]{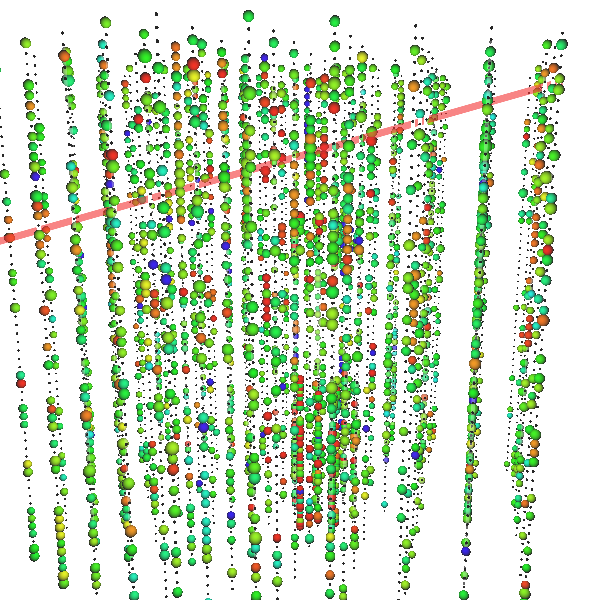}
\caption{Simulated signal event of a horizontal magnetic monopole with speed $0.01\,\textrm{c}$. The color, from red to blue, indicates the time evolution of the event. The total time of the event is $48\,\textrm{ms}$. 
The size of the spheres scales with the amount of light detected by each DOM.
The original track of the particle is shown in red. \vfill
Due to the length of the event,  a high fraction of all the DOMs of the detector have recorded noise events 
and there are also multiple coincident muons crossing the detector.}
\label{fig-8}
\end{figure}

Slow monopoles are defined to have a speed below $0.1\,\textrm{c}$.
An example simulated signal event is displayed in Fig.~\ref{fig-8}.
Due to their low speed, the event's duration is on the order of $10\,\textrm{milliseconds}$.
This duration is too long for the normal IceCube trigger and a specialized trigger, the Slow-Particle~(SLOP)~trigger~\cite{1612.05093v2} is used which selects candidate events with a rate of $13\,\textrm{Hz}$.
The SLOP trigger searches for a track of a slow moving particle in the HLC hits. This is achieved by collecting all HLC hits, dropping HLC hits which are in close temporal coincidence to reduce muon background, and then finding triplets of HLC hits which are compatible with a track of a slow moving particle.\\
Background simulation for these kind of events is not possible at the moment because of a mismatch between the simulation and actual measured background at long time scales.
Instead, data taken with the aforementioned trigger is used as a source of background.\\ 
The recorded hits are subjected to a Kalman filter~\cite{Kalman:1960-3} which separates out noise hits from the actual hits produced by the monopole. After these steps, straight cuts on multiple variables are applied to reduce the amount of data. In the last step, a Boosted Decision Tree (BDT) is utilized for the final event selection.

\subsection{Outlook}
\label{sec-outlook}
IceCube's construction was finished in 2010, reaching its full fiducial volume of $1\,\textrm{km}^3$ of ice.
All previous analyses in the concerned speed regions were performed while IceCube was in construction, so from the increase of size alone
it is expected that the slow monopole and highly relativistic monopole analyses will have discovery potential in their respective speed ranges. While the same is expected of the Low relativistic Monopole analysis, which has not yet been done in IceCube, it will also close the mass gap between experiments looking for intermediate mass monopoles and light monopoles.
\bibliography{sources}
\end{document}